\documentclass[3p,8pt]{elsarticle}
\usepackage{graphicx}
\usepackage{amsmath,amssymb}
\usepackage{amsthm}
\usepackage{arydshln}
\usepackage{rotate,rotating}
\usepackage{epsf,epic,eepic}
\usepackage{bmpsize}
\usepackage{multirow} %by sasa
\usepackage{bm}
\usepackage{mathrsfs}
\usepackage{color}

\newcommand{\mc}[1]{\mathcal{#1}}

\newtheorem{theorem}{Theorem}

\theoremstyle{remark}

\usepackage{amssymb}
\journal{ArXiv}

\begin{document}

\begin{frontmatter}



\title{Global stability of the multi-strain Kermack-McKendrick (renewal) epidemic model}

 \author[AITHM]{Michael T. Meehan}
\author[Daniel]{Daniel G. Cocks}
%\author[Johannes]{\HL {Johannes M{\"u}ller}}
\author[AITHM]{Emma S. McBryde}

\address[AITHM]{Australian Institute of Tropical Health and Medicine, James Cook University, Townsville, Australia}
\address[Daniel]{Research School of Science and Engineering, Australian National University, Canberra, Australia}
%\address[Johannes]{\HL {Centre for Mathematical Sciences, Technische Universt{\"a}t M{\"u}nchen, and Institute of Computational Biology, German Research Center for Environmental Health, M{\"u}nchen}}

%\author{Michael T. Meehan        \and
%				Daniel G. Cocks \and
%				James M. Trauer \and
%				Emma S. McBryde \and
%}

%\address{}

%\address{M. T. Meehan, E. S. McBryde \at
%              Australian Institute of Tropical Health and Medicine, James Cook University \\
  %            \email{michael.meehan1@jcu.edu.au}           %  \\
  %         \and
   %        D. G. Cocks \at
    %          College of Science and Engineering, James Cook University 
	%						\and
	%						J. M. Trauer \at
	%						School of Public Health and Preventive Medicine, Monash University
%}

\begin{abstract}
We extend a recent investigation by Meehan et al. (2019) \citep{meehan2019global} regarding the global stability properties of the general Kermack-McKendrick (renewal) model to the multi-strain case. We demonstrate that the basic reproduction number of each strain $R_{0j}$ represents a sharp threshold parameter such that when $R_{0j} \leq 1$ for all $j$ each strain dies out and the infection-free equilibrium is globally asymptotically stable, whereas for $R_{01} \equiv \mathrm{max}_j\, R_{0j} > 1$ the endemic equilibrium point $\bar{P}^1$, at which only the fittest strain (i.e. strain 1) remains in circulation, becomes globally asymptotically stable.

\end{abstract}


\end{frontmatter}






\section{Introduction}
\label{sec:Introduction}

%Why are you looking at multi-strain epidemic models?
%Many infectious diseases now have several genetic variants circulating within populations. Further stratification is required.
%We want to investigate the dynamics of systems which contain multiple circulating pathogens. This is an attempt to model ecological dynamics among several genetic (e.g. drug-resistant) variants of a particular disease. Multi-strain epidemic models are a powerful tool to investigate multi-strain dynamics.

The combination of genetic evolution and primary transmission has driven an explosion in the number of phenotypically distinct lineages of infectious diseases (i.e. strains) circulating in the global population. To simulate the dynamics of several co-circulating pathogen strains, several authors have developed multi-strain extensions of canonical single-strain epidemic models~\cite{anderson1982coevolution,Beck1984,BremermannThieme,andreasen1995pathogen,LIPSITCH199731,Ackleh2003,Bichara2013}. Often, these models are constructed by making $n$ copies of the various infectious states considered for a single strain, with the additional constraint that each of the $n$ strains dips from a common susceptible pool. In this case, many authors (e.g.~\cite{anderson1982coevolution,Beck1984,BremermannThieme}) have rediscovered the well-known competitive exclusion principle (first appearing in the ecological literature~\cite{volterra1928variations,levin1970community}) which asserts that when several species are competing over a shared resource only one of them can survive indefinitely --- namely, the one with the greatest reproductive capacity. Although this result can often be deduced by investigating the asymptotic dynamics of each system, the global stability of the various equilibria has often proven to be more difficult to establish (see for instance~\cite{BremermannThieme}). %We also point out, that several notable exceptions, i.e. models that promote strain diversity at equilibrium, have also been found~\cite{MAY199495,andreasen1995pathogen,lipsitch1996evolution,martcheva2009non,meehan2016coupled}.

Many of the multi-strain models investigated thus far have been of compartmental type for which infected individuals have a fixed infectiousness for the duration of their infectious period (which may or may not follow an exponentially distributed latency period). In this article we extend these approaches by adapting the general Kermack-McKendrick renewal model to the multi-strain context (section~\ref{sec:model}) and apply the direct Lyapunov method to establish the necessary and sufficient conditions for the global asymptotic stability of the infection-free and endemic equilibrium points of our system (section~\ref{sec:stability}). Importantly, since the general Kermack-McKendrick model incorporates many of the familiar transmission dynamic models as limiting cases (e.g. the SIR and SEIR models)~\citep{DIEKMANN1977459, metz1986,breda2012formulation}, the analysis and results presented in this article generalize a number of results derived in previous investigations~\cite{Bichara2013}. (For applications of the direct Lyapunov method to single-strain compartmental epidemic models see e.g.~\citep{Korobeinikov2002955,li1995global,fan2001global} and for renewal-type models see e.g.~\citep{magal2010lyapunov,McCluskey201055}.)

\section{Model description}
\label{sec:model}

The model we investigate in this article is a multi-strain extension of the general Kermack-McKendrick model outlined in detail in \citep{meehan2019global} (see also~\citep{kermack1927contribution, magal2010lyapunov,breda2012formulation}). In particular, we develop a model with $n$ distinct, uncoupled pathogen strains which each dip from a common susceptible pool, S. We assume perfect cross-immunity such that, once infected with strain $j\in[1,n]$, individuals are immune to further infection with an alternate strain. In the absence of cross-immunity it is possible that chaotic solutions may arise~\citep{Aguiar2008epidemiology,Minayev2008improving}.%Further details beyond the fundamental model ingredients described below can be found in \citep{meehan2019global}.

%We assume perfect cross-immunity between strains and that, once infected, individuals before recovering to state R, or dying from natural or infection-related causes.
%and enter the recovered state R upon recovery.

% For brevity however, we describe only the fundamental model ingredients and encourage the reader to consult \citep{meehan2019global} for further details.

Firstly, we introduce the force of infection of each strain $j$, $F_j(t)$, which, by definition, is the per-capita rate at which susceptibles are infected with strain $j$ at time $t$. It follows then, that the incidence of strain $j$ at time $t$, which we denote $v_j(t)$, is given by
\begin{equation}
v_j(t) = F_j(t)S(t) \nonumber% = S(t)\int_0^\infty A(\tau) F(t-\tau) S(t-\tau) \, d\tau.
\end{equation}
where $S(t)$ denotes the number of susceptibles. Here, we assume that the force of infection depends linearly on the size of each infectious population, such that $F_j(t)$ can be expressed in terms of a renewal equation:
\begin{equation}
F_j(t) = \int_0^{\bar{\tau}_j} A_j(\tau) F_j(t-\tau) S(t-\tau) \, d\tau.\label{eq:F}
%F_j(t) = \int_0^\infty A_j(\tau) F_j(t-\tau) S(t-\tau) \, d\tau.\nonumber
\end{equation}
The kernel $A_j(\tau)$ gives the expected contribution to the force of infection for an individual who has been infected with strain $j$ for $\tau$ units of time and $\bar{\tau}_j$ is the maximum infection-age at which an individual infected with strain $j$ contributes to the force of infection $F_j(t)$:
\begin{equation}
\bar{\tau}_j = \mathrm{sup}\left\{\tau\geq 0\: :\: A_j(\tau) > 0\right\} < \infty.\nonumber%\label{eq:taubar}
\end{equation}
Note that in our analysis we assume $\bar\tau_j < \infty$ for all $j$ because the lack of compactness in the infinite case makes the problem much more difficult~\citep{doi:10.1137/060659211,DIEKMANN2012819}.

We also define
\begin{equation}
\bar\tau = \max_j \bar\tau_j.\nonumber
\end{equation}

% Among these we also single out $\bar\tau = \mathrm{max}_j\bar\tau_j$. In this case, the expression for $F_j(t)$ reduces to
%\begin{equation}
%F_j(t) = \int_0^{\bar{\tau}_j} A_j(\tau) F_j(t-\tau) S(t-\tau) \, d\tau.\label{eq:F}
%\end{equation}

Next, we assume that individuals are recruited (i.e. born) directly into the susceptible class at a constant rate $\lambda$ and that all individuals experience a constant per-capita natural death rate, $\mu$. Therefore, if we combine the demographic influences with the loss of susceptible individuals due to all types of infection, we find that the susceptible population varies according to
\begin{equation}
\frac{dS(t)}{dt} = \lambda - \mu S(t) - \sum_{j = 1}^n F_j(t)S(t).\label{eq:dS}
\end{equation}
Finally, if we integrate the expected contribution to the force of infection $A_j(\tau)$ over all possible infection ages we find that the basic reproduction number for each strain, $R_{0j}$, is given by
\begin{equation}
R_{0j} = S^0\int_0^{\bar{\tau}_j} A_j(\tau)\,d\tau\label{eq:R0}
\end{equation}
where $S^0=\lambda/\mu$ is the steady-state susceptible population in the absence of infection. In the analysis that follows, we assume that 
\begin{equation}
R_{0i} \neq R_{0j} \quad \mbox{for}\quad  i\neq j, \nonumber% (see section~\ref{sec:equilibrium}). 
\end{equation}
and, without loss of generality, we label the strain with the maximum reproduction number strain~1, such that
\begin{equation}
R_{01} \equiv \max_{j} \, R_{0j}.\nonumber
\end{equation}
Supplementing our model with appropriate initial conditions, $\mc{S}_0\in C^0_+([-\bar\tau,0])$ and $\mc{F}_{j,0}\in L^1_+(-\bar\tau_j,0)$, where $\mc{S}_0$ and $\mc{F}_{j,0}$ are the initial histories of the susceptible population and the force of infection of each strain $j$ respectively, we see that the phase-space of our system is given by the product topology
\begin{equation}
\Omega = C^0_+([-\bar\tau,0])\times \prod_{j=1}^n L^1_+(-\bar\tau_j, 0)\nonumber
%\Omega = \mathbb{R}_{\geq 0} \times \prod_{j=1}^n L^1_+(t-\bar\tau,t).
\end{equation}
which is a Banach space with the natural norm
\begin{equation}
\| (\mathcal{S}, \mathcal{F}_1, \mathcal{F}_2,\ldots, \mathcal{F}_n)\| = \sup_{s\in \left[-\bar\tau,0\right]}|\mathcal{S}(s)| + \sum_{j=1}^n\int_{-\bar\tau_j}^0 |\mathcal{F}_j(s)|ds.\nonumber
\end{equation}
%In this instance, $(\mathcal{S}, \mathcal{F}_1, \mathcal{F}_2,\ldots, \mathcal{F}_n)$ is a general state in $\Omega$ (not necessarily a state along the system trajectory).

With this choice of state space standard arguments show that the model~\eqref{eq:F}-\eqref{eq:dS} is well defined and induces a continuous semi-flow $\Phi_t : \Omega\rightarrow \Omega$. 
Importantly, by Lemma 1 of~\citep{meehan2019global}, which invokes the smoothing properties of convolution integrals detailed in~\citep{Mikusinski}, it is straightforward to show that when the infectivity kernels $A_j$ are of bounded variation, i.e. $A_j\in BV([0,\bar{\tau}_j])$, system trajectories generated by the continuous semiflow $\Phi_t$ originating in $\Omega$ enter a bounded subset $\Omega^c\subset \Omega$ that is relatively compact. In this case, which we shall assume holds forthwith, the $\omega$-limit set of the system~\eqref{eq:F}-\eqref{eq:dS} is non-empty such that the infinite-dimensional form of LaSalle's extension to Lyapunov's global asymptotic stability theorem~\citep[Theorem~5.17]{hsmith_textbook} can be applied.

As a useful shorthand, we introduce the notation $\mathscr{F} = (\mc{F}_1, \mc{F}_2, \ldots)$ to denote the set of force of infection states and observe that the system trajectory $(\mc{S}_t(\cdot), \mathscr{F}_t(\cdot))\in\Omega$ with
%If we introduce the notation $\mathscr{F} = (\mc{F}_1, \mc{F}_2, \ldots)$, the system trajectory is given by $(\mc S_t, \mathscr{F}_t)\in\Omega$ where
\begin{equation}
\mc S_t(s) = S(t+s),\qquad \mathscr{F}_t(s) = \left(\mc{F}_{1,t}(s), \mc{F}_{2,t}(s), \ldots\right) = \left(F_1(t+s), F_2(t+s),\ldots\right), \qquad s\in[-\bar\tau,0].\nonumber
\end{equation}

Next, we discuss the equilibrium states of the system. We have already observed from equations~\eqref{eq:F}-\eqref{eq:dS} that the trivial infection-free equilibrium solution $P^0$ is given by $\mathscr{F}^0 = 0$ and $\mc S^0 = S^0 = \lambda/\mu$, where, in general, $P^0$ belongs to the collection of infection-free states $\partial\Omega$:
\begin{equation}
P^0 \in \partial\Omega = \left\{\left(\mc{S}, \mathscr{F}\right)\in \Omega \, | \, \mathscr{F} = 0\right\}.\nonumber
\end{equation}
%From the model equations~\eqref{eq:F}-\eqref{eq:dS} we observe that the trivial, infection-free equilibrium solution $P^0$ is given by $\mathscr{F}^0 = 0$ and $\mc S^0 = S^0 = \lambda/\mu$. 
Here we use the notation $\partial\Omega$ not in the strict topological sense, but rather in view of our application: $\partial\Omega$ is the set of states for which the force of infection is vanishing and which therefore lead to trivial dynamics.

To determine the remaining (endemic) solutions, $\bar P^j = (\bar{\mc S}, \bar{\mathscr{F}}^j)$, we solve equation~\eqref{eq:F} to obtain
\begin{equation}
\mc K[\bar{\mathscr{F}}] = \frac{\bar S}{S^0}\,\bar{\mathscr{F}}\nonumber
\end{equation}
where we introduce the next-generation operator $\mathcal{K}$, defined as 
\begin{equation}
\mathcal{K} = \mathrm{diag}\left(S^0\int_0^{\bar\tau_j} A_j(\tau)\,d\tau\right) = \mathrm{diag}\left(R_{0j}\right).\nonumber
\end{equation}
%Solving~\eqref{eq:F} at equilibrium we then get
Hence, the endemic solutions, $\bar P^j$, are determined by the spectrum of the (diagonal) operator $\mc K$:
\begin{equation}
\bar S^j = \frac{S^0}{R_{0j}} \qquad \mbox{and} \qquad \bar F^j_i = \begin{cases}
\mu(R_{0j} - 1), \quad &i = j \\
 0, &\mbox{otherwise}
\end{cases}
\label{eq:Pbar}
%\bar{\mathscr{F}} = \left(0,0,\ldots, \mu(R_{0j} - 1), \ldots, 0\right),\label{eq:Pbar}
\end{equation}
where the $\bar{\mathscr{F}}^j$ have been calculated by substituting the solution for $\bar S^j$ into~\eqref{eq:dS} and rearranging. 

From~\eqref{eq:Pbar}, we see that at each of the endemic equilibrium points $\bar{P}^j$ only strain $j$ survives, i.e. we have competitive exclusion. Moreover, $\bar P^j\in\Omega $ if, and only if $R_{0j} \geq 1$. %; for the limiting case $R_{0j} = 1$, $\bar{P}^j$ and $P^0$ coincide.

%This of course also applies to any strains that are not present initially. Active strains: That is, the set of strains that have a non-zero force of infection, either at the current time, or at some point in the future.
%who go extinct, or at least functionally so,

%\begin{equation}
%\mc A = \left\{j \in [1, n] \: | \:  \int_0^{\bar\tau}\int_0^{\bar\tau} A_j(\tau + a)\mc F_j(-\tau) \mc S(-\tau) \,d\tau\,da > 0\right\}
%\end{equation}

%As in~\citep{meehan2019global}, with this choice of state space standard arguments show that the model is well defined and that extensions generated by~\eqref{eq:F}-\eqref{eq:dS} are bounded and continuous. Moreover, we can show that system trajectories $\mc{S}_t\in C^1([-\bar\tau,0])$ and $\mc{F}_{j,t}\in W^{1,1}(-\bar\tau_j,0)$ for all $j$ when $t > \bar\tau$, such that we eventually have compactness. Therefore, since the trajectory is bounded, the $\omega$-limit set of the system~\eqref{eq:F}-\eqref{eq:dS} is non-empty.

In the following, we would like to consider the trajectories of the system~\eqref{eq:F}-\eqref{eq:dS} originating from a space for which all strains $j\in [1,n]$ are active --- that is, the set of states, $\widehat\Omega$, for which all strains have a non-zero force of infection at some point over the interval $[0,\bar\tau]$. More formally, if we define the set of states for which a subset $\mc A \subseteq \{1,\ldots,n\}$ of strains are active as
\begin{equation}
\widehat \Omega^{\mc A} = \left\{ (\mc S, \mathscr F)\in \Omega \: \left| \: \int_0^{\bar\tau_j}\int_0^{\bar\tau_j} A_j(\tau + a)\mc F_j(-\tau) \mc S(-\tau) \,d\tau\,da > 0  \iff \: j\in \mc A\right.\right\},\nonumber
\end{equation}
we then have 
\begin{equation}
\widehat\Omega = \widehat\Omega^{\{1,\ldots,n\}}.\nonumber
\end{equation}
Moreover, we see that each of the endemic equilibrium points $\bar P^j \in \widehat\Omega^j$ whilst, conversely, $\partial\Omega \subset \widehat\Omega^\varnothing$. 

Given the parallel structure of the strains in our model, it is clear that once a strain becomes inactive (i.e. $j\notin \mc A$), it remains so thereafter. Consequently, the number of active strains can only diminish as the system evolves: $\Phi_{t>0} (\widehat\Omega^{\mc A} )\subseteq \widehat\Omega^{\mc A}$. Indeed, as we will show below, in the case of endemic infection, the system evolves towards a state in which only the fittest strain (strain 1) survives, i.e. $\Phi_{t\rightarrow\infty}(\widehat\Omega^{\{1,\ldots,n\}}) \subseteq \widehat\Omega^1$.

\section{Global stability analysis}
\label{sec:stability}
% Having found the equilibrium solutions of the system~\eqref{eq:dS} and~\eqref{eq:I} we now establish their global stability properties.
We now establish the stability properties of the equilibrium solutions of the system~\eqref{eq:F}-\eqref{eq:dS}, for which we recall the definition $R_{01} \equiv \max_j R_{0j}$ to identify strain 1.

 %Since the analysis below follows that presented in \citep{meehan2019global}, we omit some of the detailed working and simply provide the necessary steps required to establish our argument.

\subsection{Infection-free equilibrium}
\begin{theorem}
\label{the:ife}
The infection-free equilibrium point $P^0$ is globally asymptotically stable in $\Omega$ if $R_{01} \leq 1$. However, if $R_{01} > 1$, solutions of~\eqref{eq:F}-\eqref{eq:dS} starting sufficiently close to $P^0$ in $\widehat\Omega = \widehat\Omega^{\{1,\ldots,n\}}$ move away from $P^0$.\footnote{The extension to systems for which only a subset of strains are active initially, i.e. $\widehat\Omega^{\mc A}$ where $\mc A \subseteq \{1,\ldots,n\}$, is trivial, provided $R_{01}$ is redefined appropriately: $R_{01} = \max_{j\in \mc A} R_{0j}$.} In all cases, solutions starting in $\partial\Omega \subset \widehat \Omega^\varnothing$ approach $P^0$.
\end{theorem}

\begin{proof}
Recall that $\bar\tau=\max_j \bar\tau_j$ and consider the forward-invariant and attracting region $D =\Phi_{\bar\tau}(\Omega)$. From~\eqref{eq:dS} we observe $\mc S(0) > 0$ for all $(\mc S, \mathscr{F})\in D$.
Next we define $U\, : \,  D\rightarrow \mathbb{R}_{+}$ (c.f.~\citep{meehan2019global}):
\begin{equation}
U(\mc S,\mathscr{F}) = g\left(\frac{\mc{S}(0)}{S^0}\right) + \sum_{j=1}^n \int_0^{\bar\tau_j} \eta_j(\tau)\mc{F}_j(-\tau)\mc{S}(-\tau)\,d\tau\label{eq:U}
\end{equation}
where
%\begin{equation}
%U_1(\mc S,\mathscr{F})  = S - S_0 \log S \qquad \mbox{and} \qquad U_2 = \sum_{j = 1}^n \int_0^{\bar{\tau}_j} \eta_j(\tau) v_j(t-\tau)\,d\tau\label{eq:U}
%\end{equation}
%and 
\begin{equation}
g(x) = x - 1 - \log{x} \quad \mbox{and} \quad \eta_j(\tau) = \int_\tau^{\bar{\tau}_j} A_j(s)\,ds.\label{eq:gdef}
\end{equation}
In particular we have $\eta_j(\bar{\tau}_j) = 0$,
\begin{equation}
\eta_j(0) = \frac{R_{0j}}{S^0} \qquad \mbox{and} \qquad \eta'_j(\tau) = -A_j(\tau) \label{eq:etacons}
\end{equation}
where a $'$ denotes differentiation with respect to $\tau$. Note, the functional $U$ is positive, continuous and well defined in $D$, and has a global minimum in $\Omega$ at $P^0$.

Using $\mc{S}_t(s) = S(t+s)$ and $\mc{F}_{j,t}(s) = F_j(t+s)$ and rewriting the integral in~\eqref{eq:U} we find that $U$, evaluated along system trajectories, is given by
\begin{align}
U(\mc S_t(\cdot),\mathscr{F}_t(\cdot)) &= g\left(\frac{\mc{S}_t(0)}{S^0}\right) + \sum_{j=1}^n \int_0^{\bar\tau_j} \eta_j(\tau)\mc{F}_{j,t}(-\tau)\mc{S}_t(-\tau)\,d\tau,\nonumber\\
%&= g\left(\frac{S(t)}{S^0}\right) + \sum_{j=1}^n \int_0^{\bar\tau_j} \eta_j(\tau) F_{j}(t-\tau)S(t-\tau)\,d\tau,\nonumber\\
&= g\left(\frac{S(t)}{S^0}\right) + \sum_{j=1}^n \int_{t-\bar\tau_j}^{t} \eta_j(t-s)F_{j}(s)S(s)\,ds.\label{eq:Utrans}
\end{align}
Taking the time derivative of each term in~\eqref{eq:Utrans} separately, and substituting in the model equations~\eqref{eq:F} and~\eqref{eq:dS}, we get % (see \citep{meehan2019global} for further details)
\begin{align}
\frac{d}{dt}\,g\left(\frac{S(t)}{S^0}\right) &= \left(\frac{1}{S^0} - \frac{1}{S(t)}\right)\,\frac{dS(t)}{dt},\nonumber\\
&= \frac{\lambda}{S^0} - \mu\,\frac{S(t)}{S^0} - \sum_{j=1}^n F_j(t)\,\frac{S(t)}{S^0} - \frac{\lambda}{S(t)} + \mu + \sum_{j=1}^n F_j(t),\nonumber\\
&= \mu\left(2 - \frac{S(t)}{S^0} - \frac{S^0}{S(t)}\right) - \sum_{j=1}^n F_j(t)\left(\frac{S(t)}{S^0} - 1\right),\nonumber\\
&= -\mu\,\frac{S(t)}{S^0}\left(1 - \frac{S^0}{S(t)}\right)^2 - \sum_{j=1}^n F_j(t)\left(\frac{S(t)}{S^0} - 1\right)\nonumber
\end{align}
where in the second line we have substituted in the identity $\lambda = \mu S^0$, and
\begin{align}
\frac{d}{dt} \left[\sum_{j=1}^n \int_{t-\bar\tau_j}^{t}\eta_j(t-s)F_j(s)S(s)\,ds\right] &= \sum_{j=1}^n \Bigg[\eta_j(0)F_j(t)S(t) - \eta_j(\bar\tau_j)F_j(t-\bar\tau_j)S(t-\bar\tau_j)\nonumber\\
&\qquad \qquad + \int_{t-\bar\tau_j}^{t} \eta'(t-s) F_j(s)S(s)\,ds\Bigg],\nonumber\\
&= \sum_{j=1}^n \left[R_{0j}F_j(t)\frac{S(t)}{S^0} - \int_{t-\bar\tau_j}^t A_j(t-s)F_j(s)S(s)\,ds\right],\nonumber\\
&= \sum_{j=1}^n F_j(t)\left(R_{0j}\,\frac{S(t)}{S^0} - 1\right).\nonumber
\end{align}
%Calculating the time derivative of~\eqref{eq:U} along the system trajectories, we find (see \citep{meehan2019global} for further details)
Combining these results we then have
\begin{align}
\frac{d}{dt}U(\mc{S}_t,\mathscr{F}_t) &= -\mu \frac{\mc{S}_t(0)}{S^0}\left(1 - \frac{S^0}{\mc{S}_t(0)}\right)^2 -\sum_{j  = 1}^n \left(1 - R_{0j}\right)\mc{F}_{j,t}(0)\,\frac{\mc{S}_t(0)}{S^0} \quad\leq 0.\label{eq:dU}
\end{align}
%Given the conditions cited above we know that this expression is well defined and $U$ is a proper functional on $D$. Moreover, 
For system trajectories $(\mc{S}_t, \mathscr{F}_t)\in D \subset \Omega$ equations~\eqref{eq:F} and~\eqref{eq:dS} imply that $\mc{F}_{j,t} \in C^0([-\bar\tau_j, 0])$ for $t > \bar\tau$ (see~\citep{meehan2019global}) such that~\eqref{eq:dU} is well defined and $U$ is a proper Lyapunov functional on the domain $D$.

We observe that the derivative $\dot{U} = 0$ if and only if $\mc S_t(0) = S^0$ and either (a) $R_{0j} = 1$  or (b) $\mc F_{j,t}(0) = 0$ for all $j$. Therefore, the largest invariant subset in $\Omega$ for which $\dot{U} = 0$ is the singleton $\left\{P^0\right\}$. Since by Lemma 1 of~\citep{meehan2019global} the orbit is eventually precompact, by the infinite-dimensional form of LaSalle's extension of Lyapunov's global asymptotic stability theorem~\citep[Theorem~5.17]{hsmith_textbook}, the infection-free equilibrium point $P^0$ is globally asymptotically stable in $\Omega$ if $R_{01} \leq 1$.

Conversely, if $R_{0j} > 1$ for any $j$, the derivative $\dot{U} > 0$ for $\mc{S}_t(0)$ sufficiently close to $S^0$, provided $\mc{F}_{j,t}(0) > 0$. Therefore, solutions starting sufficiently close to the infection-free equilibrium point $P^0$ in $\widehat\Omega^j$ leave a neighbourhood of $P^0$. Since $\dot{U} \leq 0$ for solutions starting in $\partial\Omega$ these solutions approach $P^0$.

\end{proof}

\subsection{Endemic equilibrium}
For convenience, in this section we adopt the shorthand notation that an overbar refers to the value of a state variable at the endemic equilibrium point $\bar{P}^1$ such that, for example, $\bar{\mc S} \equiv \bar{\mc S}^1$ and $\bar{\mathscr F} \equiv  \bar{\mathscr F}^1$.

\begin{theorem}
If $R_{01} > 1$ the endemic equilibrium point $\bar{P}^1$ is globally asymptotically stable in $\widehat\Omega = \widehat\Omega^{\{1,\ldots,n\}}$.\footnote{As before, the extension to systems for which only a subset of strains are active initially, i.e. $\widehat\Omega^{\mc A}$ where $\mc A \subseteq \{1,\ldots,n\}$, is trivial, provided $R_{01}$ is redefined appropriately: $R_{01} = \max_{j\in \mc A} R_{0j}$.}
\end{theorem}

\begin{proof}
From theorem~\ref{the:ife} we have that $F_1(t)$ is bounded away from zero for $t > 0$ when $R_{01} > 1$, such that $\Phi_t : \widehat{\Omega}\rightarrow \widehat{\Omega}^{\mc A}\supseteq \widehat\Omega^1$.\footnote{Although $F_1(t)$ is not uniformly bounded away from zero, it is sufficiently so to ensure that our Lyapunov function $W$ remains well defined.} Recall that $\bar\tau=\max_j\bar\tau_j$ and define $\widehat D = \Phi_{\bar\tau}(\widehat{\Omega})$ which is forward invariant and attracting for $R_{01} > 1$. Moreover, $\mc{S}, \mc F_1 > 0$ for $(\mc S, \mathscr{F})\in \widehat{D}$.
Consider the Lyapunov functional $W\, : \,  \widehat D\rightarrow \mathbb{R}_+$ (c.f.~\citep{meehan2019global}):
\begin{equation}
%W = W_1 + W_2 + W_3\label{eq:W}
W(\mc S, \mathscr{F}) = g\left(\frac{\mc{S}(0)}{\bar S}\right) + \bar F_1 \bar S \int_0^{\bar\tau_1} \chi_1(\tau) g\left(\frac{\mc{F}_1(-\tau)\mc S(-\tau)}{\bar F_1 \bar S}\right) d\tau + \sum_{j = 2}^n \int_0^{\bar\tau_j} \chi_j(\tau) \mc F_{j}(-\tau) \mc{S}(-\tau)d\tau \label{eq:W_init}
\end{equation}
where $g(x)$ has been defined previously in~\eqref{eq:gdef} and
%\begin{align}
%W_1 &= S - \bar{S}\log S,\nonumber\\
%W_2 &= \sum_{j = 1}^n\int_0^{\bar{\tau}_j} \chi_j(\tau)v_j(t-\tau)\,d\tau,\nonumber\\
%W_3 &= -\bar{v}\int_0^{\bar{\tau}_1} \chi_1(\tau) \log v_1(t-\tau)\,d\tau\nonumber
%\end{align}
%with
\begin{equation}
\chi_j(\tau) = \int_\tau^{\bar{\tau}_j} A_j(s)\,ds.\nonumber
\end{equation}
In particular, we have
\begin{equation}
\chi_j(0) = \frac{R_{0j}}{S^0} = \frac{R_{0j}}{R_{01}} \,\frac{1}{\bar S}\qquad \mbox{and} \qquad \chi_j'(\tau) = -A_j(\tau)\label{eq:chiprop}
\end{equation}
so that $\chi_1(0) = 1/\bar S$ and $\chi_j(\bar{\tau}_j) = 0$. Note the functional $W$ is positive, continuous, and has a global minimum in $\widehat{\Omega}$ at $\bar{P}^1$.

Rewriting the integral terms as in theorem~\ref{the:ife} and evaluating $W$ along system trajectories gives
\begin{equation}
W(\mc S_t, \mathscr{F}_t) = g\left(\frac{S(t)}{\bar S}\right) + \bar F_1 \bar S \int_{t-\bar\tau_1}^{t} \chi_1(t-s) g\left(\frac{F_1(s) S(s)}{\bar F_1 \bar S}\right) ds + \sum_{j = 2}^n \int_{t-\bar\tau_j}^{t} \chi_j(t-s) F_{j}(s) S(s)ds. \label{eq:W}
\end{equation}

Next we differentiate the first term in~\eqref{eq:W}  to get %(see \citep{meehan2019global} for details)
\begin{align}
\frac{d}{dt} g\left(\frac{S(t)}{\bar S}\right) &= \left(\frac{1}{\bar S} - \frac{1}{S(t)}\right)\,\frac{dS(t)}{dt},\nonumber\\
&= \frac{\lambda}{\bar S} - \mu \,\frac{S(t)}{\bar S} - \sum_{j = 1}^n F_j(t)\,\frac{S(t)}{\bar S} - \frac{\lambda}{S(t)} + \mu +\sum_{j = 1}^n F_j(t),\nonumber\\
&= \mu\left(2 - \frac{S(t)}{\bar S} - \frac{\bar S}{S(t)}\right) + \sum_{j=1}^n \bar F_j\left(1 - \frac{\bar S}{S(t)}\right) + \sum_{j=1}^n F_j(t)\left(1 - \frac{S(t)}{\bar S}\right),\nonumber\\
&= -\mu \frac{S(t)}{\bar{S}}\left(1 - \frac{\bar S}{S(t)}\right)^2 + \bar{F}_1\left(1 - \frac{\bar S}{S(t)}\right) + \sum_{j=1}^n F_j(t)\left(1 - \frac{S(t)}{\bar S}\right),\label{eq:W1dot}
\end{align}
where in the second line we have used the identity $\lambda = \mu \bar S + \sum_{j=1}^n \bar F_j \bar S = \mu \bar S + \bar F_1 \bar S$. Similarly, we differentiate the final term in~\eqref{eq:W} and substitute in the definition of $\chi_j(\tau)$ to get
\begin{align}
\frac{d}{dt} \left[\sum_{j = 2}^n \int_{t-\bar\tau_j}^{t} \chi_j(t-s) F_{j}(s) S(s)ds\right] &= \sum_{j=2}^n \Bigg[\chi_j(0)F_j(t)S(t) - \chi_j(\bar\tau_j)F_j(t-\bar\tau_j)S(t-\bar\tau_j)\nonumber\\
&\qquad \qquad + \int_{t-\bar\tau_j}^t \chi_j'(t-s)F_j(s)S(s)\,ds\Bigg],\nonumber\\
&= \sum_{j=2}^n\left[F_j(t)\,\frac{R_{0j}}{R_{01}}\frac{S(t)}{\bar S} - \int_{t-\bar\tau_j}^t A_j(t-s)F_j(s)S(s)\,ds\right],\nonumber\\
&=\sum_{j=2}^n F_j(t)\left(\frac{R_{0j}}{R_{01}}\frac{S(t)}{\bar{S}} - 1\right).\label{eq:W2dot}
\end{align}
Differentiating the second term in~\eqref{eq:W} and using~\eqref{eq:chiprop}, yields
\begin{align}
\frac{d}{dt} \left[\bar{F}_1\bar S\int_{t-\bar\tau_1}^t \chi_1(t-s) g\left(\frac{F_1(s) S(s)}{\bar F_1\bar S}\right) ds\right] &= \bar F_1 \bar S\left[\chi_1(0) g\left(\frac{F_1(t) S(t)}{\bar F_1 \bar S}\right) - \chi_1(\bar\tau_1) g\left(\frac{F_1(t-\bar\tau_1) S(t-\bar\tau_1)}{\bar F_1 \bar S}\right)\right.\nonumber\\
&\qquad \qquad + \left. \int_{t-\bar\tau_1}^t \chi_1'(t-s) g\left(\frac{F_1(s) S(s)}{\bar F_1 \bar S}\right) ds\right],\nonumber\\
&= \bar{F}_1 g\left(\frac{F_1(t) S(t)}{\bar F_1 \bar S}\right)  - \bar{F}_1\bar{S}\int_{t-\bar\tau_1}^t A_1(t-s) g\left(\frac{F_1(s) S(s)}{\bar F_1 \bar S}\right)\,ds,\nonumber\\
&= F_1(t)\left(\frac{S(t)}{\bar S} - 1\right) - \bar{F}_1\log \left(\frac{F_1(t)S(t)}{\bar F_1\bar S}\right)\nonumber\\
&\qquad \qquad + \bar F_1\bar S \int_{t-\bar\tau_1}^t A_1(t-s) \log\left(\frac{F_1(s) S(s)}{\bar F_1 \bar S}\right)ds\nonumber
\end{align}
where in the last line we have invoked the identity $\bar S\int_0^{\bar\tau_1} A_1(\tau)\,d\tau = 1$. As in \citep{meehan2019global}, we can bound this expression using Jensen's inequality:\footnote{For a concave function $\varphi(\cdot)$, and probability distribution $h(t)$, the following inequality holds:
\begin{equation}
\varphi\left(\int_0^\infty h(t)f(t) \,dt\right) \geq \int_0^\infty h(t)\varphi\left(f(t)\right)\,dt.\nonumber
\end{equation}}
\begin{align}
\frac{d}{dt} \left[\bar{F}_1\bar S\int_{t-\bar\tau_1}^t \chi_1(t-s) g\left(\frac{F_1(s) S(s)}{\bar F_1 \bar S}\right) ds\right] &\leq F_1(t)\left(\frac{S(t)}{\bar S} - 1\right) - \bar{F}_1\log \left(\frac{F_1(t)S(t)}{\bar F_1\bar S}\right)\nonumber\\
& \qquad \qquad + \bar F_1 \log\left[\bar S \int_{t-\bar\tau_1}^t A_1(t-s) \frac{F_1(s) S(s)}{\bar F_1 \bar S}ds\right],\nonumber\\
&\leq F_1(t)\left(\frac{S(t)}{\bar S} - 1\right) - \bar{F}_1\log \left(\frac{F_1(t)S(t)}{\bar F_1\bar S}\right) + \bar F_1\log \left(\frac{F_1(t)}{\bar F_1}\right),\nonumber\\
&\leq F_1(t)\left(\frac{S(t)}{\bar S} - 1\right) - \bar{F}_1\log \left(\frac{S(t)}{\bar S}\right),\nonumber\\
&\leq F_1(t)\left(\frac{S(t)}{\bar S} - 1\right) - \bar{F}_1\left(1 - \frac{\bar S}{S(t)}\right)\label{eq:W3dot}
\end{align}
where in the last line we have also used $\log x \geq 1 - \frac{1}{x}$.

%Substituting this result back into~\eqref{eq:W3dotexact} we then have
%\begin{equation}
%\frac{dW_3}{dt} \leq -\bar{F}\bar{S}\left(1 - \frac{\bar{S}}{S}\right).\label{eq:W3dotbound}
%\end{equation}

Finally, recalling that $R_{01} = \mathrm{max}_j R_{0j}$, and combining~\eqref{eq:W1dot},~\eqref{eq:W2dot} and~\eqref{eq:W3dot} yields
\begin{align}
\frac{d}{dt}W(\mc{S}_t, \mathscr{F}_t) &\leq -\mu \frac{\mc S_t(0)}{\bar S}\left(1 - \frac{\bar{S}}{\mc S_t(0)}\right)^2 - \sum_{j = 2}^n \left(1 - \frac{R_{0j}}{R_{01}}\,\right)\mc F_{j,t}(0)\frac{\mc S_t(0)}{\bar S} \: \leq 0.\label{eq:Wdot}
\end{align}
From equation~\eqref{eq:Wdot} we see that the largest invariant subset in $\widehat\Omega$ for which $\dot{W} = 0$ is the endemic equilibrium point $\bar{P}^1$. Since by Lemma 1 of~\citep{meehan2019global} the orbit is eventually precompact, by LaSalle's extension of Lyapunov's asymptotic stability theorem~\citep[Theorem~5.17]{hsmith_textbook}, the endemic equilibrium point $\bar{P}^1$ is globally asymptotically stable in $\widehat\Omega$ for $R_{01}>1$.
\end{proof}

%\begin{remark}
%The inequality~\eqref{eq:Wdot} is only satisfied globally for strain 1, i.e. the fittest strain.
%\end{remark}

\section{Conclusions}
\label{sec:conclusions}
In this article we investigated the global stability properties of the multi-strain Kermack-McKendrick model. We found that when the basic reproduction number $R_{0j} \leq 1$ for all strains $j$ the infection-free equilibrium $P^0$ is unique in $\Omega$ and is globally asymptotically stable. We also discovered a set of $n$ endemic equilibrium solutions, $\left\{\bar{P}^j\right\}$, at which only strain $j$ survives with a positive infected population, \textit{\`a la} competitive exclusion. Moreover, we found that $\bar{P}^j$ only exists in the positive cone $\widehat\Omega$ if $R_{0j} > 1$. Our main result, which was derived using the direct Lyapunov method, was to show that of this set, $\bar{P}^1$ --- at which the fittest strain, defined by $R_{01} \equiv \max_j R_{0j}$, survives indefinitely --- is globally asymptotically stable when it exists.

\section*{Acknowledgements}

The authors would like to express their sincere gratitude to Prof. Johannes M{\" u}ller for his invaluable advice and feedback in the preparation of this manuscript.

% Importantly, our analysis generalizes a number of previous investigations into the global stability of multi-strain epidemic models. 

%Secondly, we also found that at each of the $n$ endemic equilibrium points $\bar{P}^j$, which only exist in the physical region $\mathbb{R}_{\geq 0}^{2n+1}$ if $R_{0j} > 1$, only strain $j$ survives with a positive infected population, \textit{\`a la} competitive exclusion. Moreover, we showed that of the $n$ possible endemic equilibrium points $\bar{P}^j$ only $\bar{P}^1$ at which the strain with the greatest reproduction number --- defined by $R_{01} \equiv \max_j R_{0j}$ --- survives is globally asymptotically stable when it exists. To derive this result we applied the direct Lyapunov method for which we identified appropriate Lyapunov functionals. Importantly, our analysis generalizes a number of previous investigations into the global stability of multi-strain epidemic models. 

%Importantly, our analysis conforms with the principle of competitive exclusion

%Firstly, we discussed how when the basic reproduction number $R_0 \leq 1$ the infection-free equilibrium point $P_0$ is the unique equilibrium in $\mathbb{R}^3_{\geq 0}$. In contrast, when $R_0 > 1$, an endemic equilibrium solution emerges in $\mathbb{R}_{> 0}^3$ for which a positive fraction of the population remains infected. By introducing appropriate Lyapunov functions we established that the infection-free and endemic equilibria are globally asymptotically stable when $R_0 \leq 1$ and $R_0 > 1$, respectively. 

%\bibliographystyle{elsarticle-num}  
\bibliographystyle{plainnat}

\bibliography{References}   % name your BibTeX data base

\begin{thebibliography}{25}
\providecommand{\natexlab}[1]{#1}
\providecommand{\url}[1]{\texttt{#1}}
\expandafter\ifx\csname urlstyle\endcsname\relax
  \providecommand{\doi}[1]{doi: #1}\else
  \providecommand{\doi}{doi: \begingroup \urlstyle{rm}\Url}\fi

\bibitem[Ackleh and Allen(2003)]{Ackleh2003}
S.~Ackleh and J.S. Allen.
\newblock Competitive exclusion and coexistence for pathogens in an epidemic
  model with variable population size.
\newblock \emph{Journal of Mathematical Biology}, 47\penalty0 (2):\penalty0
  153--168, 2003.
\newblock ISSN 1432-1416.
\newblock \doi{10.1007/s00285-003-0207-9}.

\bibitem[Aguiar et~al.(2008)Aguiar, Kooi, and
  Stollenwerk]{Aguiar2008epidemiology}
M.~Aguiar, B.~Kooi, and N.~Stollenwerk.
\newblock Epidemiology of dengue fever: a model with temporary cross-immunity
  and possible secondary infection shows bifurcation and chaotic behaviour in
  wide parameter regions.
\newblock \emph{Math. Model. Nat. Pheno.}, 3\penalty0 (4):\penalty0 48--70,
  January 2008.

\bibitem[Anderson and May(1982)]{anderson1982coevolution}
R.M. Anderson and R.M. May.
\newblock Coevolution of hosts and parasites.
\newblock \emph{Parasitology}, 85\penalty0 (02):\penalty0 411--426, 1982.

\bibitem[Andreasen and Pugliese(1995)]{andreasen1995pathogen}
V.~Andreasen and A.~Pugliese.
\newblock Pathogen coexistence induced by density-dependent host mortality.
\newblock \emph{Journal of Theoretical Biology}, 177\penalty0 (2):\penalty0
  159--166, 1995.

\bibitem[Beck(1984)]{Beck1984}
K.~Beck.
\newblock Coevolution: Mathematical analysis of host-parasite interactions.
\newblock \emph{Journal of Mathematical Biology}, 19\penalty0 (1):\penalty0
  63--77, 1984.
\newblock ISSN 1432-1416.
\newblock \doi{10.1007/BF00275931}.

\bibitem[Bichara et~al.(2013)Bichara, Iggidr, and Sallet]{Bichara2013}
D.~Bichara, A.~Iggidr, and G.~Sallet.
\newblock Global analysis of multi-strains {SIS}, {SIR} and {MSIR} epidemic
  models.
\newblock \emph{Journal of Applied Mathematics and Computing}, 44\penalty0
  (1):\penalty0 273--292, 2013.

\bibitem[Breda et~al.(2012)Breda, Diekmann, De~Graaf, Pugliese, and
  Vermiglio]{breda2012formulation}
D.~Breda, O.~Diekmann, W.F. De~Graaf, A.~Pugliese, and R.~Vermiglio.
\newblock On the formulation of epidemic models (an appraisal of {K}ermack and
  {M}c{K}endrick).
\newblock \emph{Journal of Biological Dynamics}, 6\penalty0 (sup2):\penalty0
  103--117, 2012.

\bibitem[Bremermann and Thieme(1989)]{BremermannThieme}
H.J. Bremermann and H.R. Thieme.
\newblock A competitive exclusion principle for pathogen virulence.
\newblock \emph{Journal of Mathematical Biology}, 27\penalty0 (2):\penalty0
  179--190, 1989.
\newblock ISSN 1432-1416.
\newblock \doi{10.1007/BF00276102}.

\bibitem[Diekmann(1977)]{DIEKMANN1977459}
O.~Diekmann.
\newblock Limiting behaviour in an epidemic model.
\newblock \emph{Nonlinear Analysis: Theory, Methods \& Applications},
  1\penalty0 (5):\penalty0 459 -- 470, 1977.
\newblock ISSN 0362-546X.

\bibitem[Diekmann and Gyllenberg(2012)]{DIEKMANN2012819}
O.~Diekmann and M.~Gyllenberg.
\newblock Equations with infinite delay: Blending the abstract and the
  concrete.
\newblock \emph{Journal of Differential Equations}, 252\penalty0 (2):\penalty0
  819 -- 851, 2012.
\newblock ISSN 0022-0396.

\bibitem[Diekmann et~al.(2008)Diekmann, Getto, and
  Gyllenberg]{doi:10.1137/060659211}
O.~Diekmann, P.~Getto, and M.~Gyllenberg.
\newblock Stability and bifurcation analysis of {V}olterra functional equations
  in the light of {S}uns and {S}tars.
\newblock \emph{SIAM Journal on Mathematical Analysis}, 39\penalty0
  (4):\penalty0 1023--1069, 2008.

\bibitem[Fan et~al.(2001)Fan, Li, and Wang]{fan2001global}
M.~Fan, M.Y. Li, and K.~Wang.
\newblock Global stability of an {SEIS} epidemic model with recruitment and a
  varying total population size.
\newblock \emph{Mathematical Biosciences}, 170\penalty0 (2):\penalty0 199--208,
  2001.

\bibitem[Kermack and McKendrick(1927)]{kermack1927contribution}
W.O. Kermack and A.G. McKendrick.
\newblock A contribution to the mathematical theory of epidemics.
\newblock \emph{Proceedings of the Royal Society of London A: Mathematical,
  Physical and Engineering Sciences}, 115\penalty0 (772):\penalty0 700--721,
  1927.

\bibitem[Korobeinikov and Wake(2002)]{Korobeinikov2002955}
A.~Korobeinikov and G.C. Wake.
\newblock Lyapunov functions and global stability for {SIR}, {SIRS}, and {SIS}
  epidemiological models.
\newblock \emph{Applied Mathematics Letters}, 15\penalty0 (8):\penalty0
  955--960, 2002.

\bibitem[Levin(1970)]{levin1970community}
S.A. Levin.
\newblock Community equilibria and stability, and an extension of the
  competitive exclusion principle.
\newblock \emph{American Naturalist}, pages 413--423, 1970.

\bibitem[Li and Muldowney(1995)]{li1995global}
M.Y. Li and J.S. Muldowney.
\newblock Global stability for the {SEIR} model in epidemiology.
\newblock \emph{Mathematical Biosciences}, 125\penalty0 (2):\penalty0 155--164,
  1995.

\bibitem[Lipsitch and Moxon(1997)]{LIPSITCH199731}
M.~Lipsitch and E.R. Moxon.
\newblock Virulence and transmissibility of pathogens: what is the
  relationship?
\newblock \emph{Trends in Microbiology}, 5\penalty0 (1):\penalty0 31 -- 37,
  1997.
\newblock ISSN 0966-842X.
\newblock \doi{http://dx.doi.org/10.1016/S0966-842X(97)81772-6}.

\bibitem[Magal et~al.(2010)Magal, McCluskey, and Webb]{magal2010lyapunov}
P.~Magal, C.C. McCluskey, and G.F. Webb.
\newblock Lyapunov functional and global asymptotic stability for an
  infection-age model.
\newblock \emph{Applicable Analysis}, 89\penalty0 (7):\penalty0 1109--1140,
  2010.

\bibitem[McCluskey(2010)]{McCluskey201055}
C.C. McCluskey.
\newblock Complete global stability for an {SIR} epidemic model with delay:
  distributed or discrete.
\newblock \emph{Nonlinear Analysis: Real World Applications}, 11\penalty0
  (1):\penalty0 55--59, 2010.

\bibitem[Meehan et~al.(2019)Meehan, Cocks, M{\"u}ller, and
  McBryde]{meehan2019global}
M.T. Meehan, D.G. Cocks, J.~M{\"u}ller, and E.S. McBryde.
\newblock Global stability properties of a class of renewal epidemic models.
\newblock \emph{Journal of Mathematical Biology}, 78:\penalty0 1713--1725,
  2019.

\bibitem[Metz and Diekmann(1986)]{metz1986}
J.~A.~J. Metz and O.~Diekmann.
\newblock \emph{The Dynamics of Physiologically Structured Populations}.
\newblock Lecture Notes in Biomathematics 68. Springer-Verlag, Berlin,
  Heidelberg, New York, London, Paris, Tokyo, 1986.

\bibitem[Mikusi{\'n}ski and Ryll-Nardzewski(1951)]{Mikusinski}
J.~Mikusi{\'n}ski and Cz. Ryll-Nardzewski.
\newblock Sur le produit de composition.
\newblock \emph{Studia Mathematica}, 12:\penalty0 51--57, 1951.

\bibitem[Minayev and Ferguson(2008)]{Minayev2008improving}
P.~Minayev and N.~Ferguson.
\newblock Improving the realism of determinstic multi-strain models:
  implications for modelling influenza a.
\newblock \emph{J. R. Soc. Interface}, 6\penalty0 (35):\penalty0 509--518,
  September 2008.

\bibitem[Smith(2010)]{hsmith_textbook}
H.~Smith.
\newblock \emph{An introduction to delay differential equations with
  applications to the life sciences}.
\newblock Springer, 2010.

\bibitem[Volterra(1928)]{volterra1928variations}
V.~Volterra.
\newblock Variations and fluctuations of the number of individuals in animal
  species living together.
\newblock \emph{J. Cons. Int. Explor. Mer}, 3\penalty0 (1):\penalty0 3--51,
  1928.

\end{thebibliography}


\end{document}